\documentclass[pra,twocolumn,preprintnumbers,superscriptaddress,longbibliography, 10pt]{revtex4-2} 

\usepackage{graphicx}
\usepackage{bm}
\usepackage{amsmath}
\usepackage{amssymb}
\usepackage{amsfonts}
\usepackage{slashed}
\usepackage{latexsym,epsfig,bbm}
\usepackage{algorithm}
\usepackage{algorithmic}
\usepackage{diagbox}

\newcommand{\bra}[1]{\langle{#1}|}
\newcommand{\ket}[1]{|{#1}\rangle}

\newcommand{\braket}[2]{\langle{#1}|{#2}\rangle}

\newcommand{\figref}[1]{Fig.~\ref{#1}}

\newcommand{\Tr}{\mathrm{Tr}}
\usepackage{color}
\definecolor{blue}{rgb}{0,0.2,1}

\definecolor{red}{rgb}{0.9,0,0}

\begin{document}

\title{Strict advantage of complex quantum theory in a communication task}

\author{Thomas J.~Elliott}
\email{physics@tjelliott.net}
\affiliation{Department of Physics \& Astronomy, University of Manchester, Manchester M13 9PL, United Kingdom}
\affiliation{Department of Mathematics, University of Manchester, Manchester M13 9PL, United Kingdom}

\date{\today}

\begin{abstract}
Standard formulations of quantum theory are based on complex numbers: Quantum states can be in superpositions, with weights given by complex probability amplitudes. Motivated by quantum theory promising a range of practical advantages over classical for a multitude of tasks, we investigate how the presence of complex amplitudes in quantum theory can yield operational advantages over counterpart real formulations. We identify a straightforward communication task for which complex quantum theory exhibits a provably lower communication cost than not just any classical approach, but also any approach based on real quantum theory. We certify the necessity of complex quantum theory for optimal approaches to the task through geometric properties of quantum state ensembles that witness the presence of basis-independent complexity. This substantiates a strict operational advantage of complex quantum theory. We discuss the relevance of this finding for quantum advantages in stochastic simulation. 
\end{abstract}
\maketitle 

\section{Introduction}

Typically, an introduction to quantum theory begins with the concept of superpositions of states, with weights given by probability amplitudes. We learn that we must take the modulus-squared of these amplitudes to determine the probabilities associated with measurement outcomes. Modulus-squared, because these amplitudes can be complex. We encounter complex numbers in many other fields of physics, but tend to think of them as merely a convenient mathematical representation, for example, to describe the phases of waves without need for an abundance of trigonometric functions. Yet with quantum theory we tend to think of complex numbers as being `physically real'~\footnote{Pun somewhat intended.}. While we can reformulate complex quantum theory in terms of real numbers alone through use of composite systems -- where, for example, a complex qubit can be embedded within a pair of real qubits (`rebits')~\cite{stueckelberg1960quantum, myrheim1999quantum, mckague2009simulating, vedral2023quantum} -- this requires an enlarged dimension of the state space, the need to impose extra dynamical constraints, and somehow appears a much less natural approach than the standard complex formulation.

A recent theoretical proposal~\cite{renou2021quantum} -- later verified in experiment~\cite{li2022testing, chen2022ruling} -- devised a means to discriminate between real and complex quantum theory. A test resemblant of Bell-like inequalities was devised, where the bounds of the associated quantity were set not under the assumption of local realism (as in a typical Bell test of quantum theory~\cite{bell1964einstein, clauser1969proposed, aspect1982experimental}), but rather, under the assumption of a real quantum theory. The violation of these bounds vindicates that (certain classes of) real quantum theory can be ruled out in favour of complex quantum theory. Here, we use the same definitions of real and complex quantum theory as Ref.~\cite{renou2021quantum}, recapitulated in Appendix \ref{sec:rcqt}.

Concurrently, a resource theory of imaginarity is being developed, seeking to quantify the degree to which a quantum state is imaginary, and to identify tasks that depend on access to this as a resource~\cite{wu2021operational, kondra2023real}. Yet, a drawback of present formulations of such a resource theory based on states is that they are explicitly basis dependent, and so states that are resourceful in one basis might not be in another. While the analogous consideration in the resource theory of coherence~\cite{baumgratz2014quantifying, yadin2017resource, streltsov2017colloquium} can often be justified when there is a preferred basis choice -- such as eigenstates of an underlying Hamiltonian, or pointer states of a particular environmental decoherence -- imaginarity is on much shakier ground in this respect.

Consider, for example, the state $(\ket{0}+i\ket{1})/\sqrt{2}$. The appearance of the imaginary amplitude here means that a resource theory of imaginarity such as that of Ref.~\cite{wu2021operational} will identify this as having non-zero imaginarity, at least with respect to the $\{\ket{0},\ket{1}\}$ basis. Yet, there is likely little -- or even no -- justification why $\ket{1}$ is chosen as a basis state, as opposed to its rescaling by an imaginary factor $\ket{1_i}:=i\ket{1}$. The above state can be written as $(\ket{0}+\ket{1_i})/\sqrt{2}$, thus exhibiting no imaginarity in the $\{\ket{0},\ket{1_i}\}$ basis. Yet, the two bases only differ in terms of an unobservable phase factor in the choice of basis states.

A proposed route towards resolving this is to consider the presence of such resources in ensembles of states, and employ geometric measures of such ensembles. For quantum states we can define geometric quantities in terms of their overlaps, such that they are basis-invariant. These can serve as witnesses to certify that coherence or imaginarity is present in an ensemble of states -- and that this cannot be removed through change of basis~\cite{oszmaniec2024measuring, fernandes2024unitary}. Such \emph{set imaginarity}~\cite{fernandes2024unitary} is a stronger condition than merely having non-real amplitudes with respect to a particular basis, as per the resource theory of imaginarity.

Here, we apply such an invariant in the context of a communication task, to be described below. The figure of merit (or `cost') in this task is the dimension of the system that must be communicated between two parties in order to successfully complete the task. For a particular instance of this task, we use a geometric invariant to certify that the minimal cost can only be achieved with complex quantum theory -- and that any approach working within the constraints of classical or real quantum theory would exhibit a higher cost. Thus, we reveal a new task for which complex quantum theory exhibits a strict operational advantage, not only over classical theory, but also over real quantum theory. This complements similar results showing a higher success probability of complex quantum theory for quantum random access codes~\cite{navascues2015characterizing}. We conclude with a discussion of further lines of enquiry into this advantage, and potential applications thereof, particularly in the context of stochastic simulation.

\section{Framework}

\subsection{Setup} 

The task under consideration can be cast in terms of a `game' with three parties. We designate one party (Charlie) as the referee, and the other two (Alice and Bob) as the two players. The game takes place over a series of rounds, of which there may be arbitrarily many. In each round, Charlie will send Alice a random variable $X$ that can take values $x\in\mathcal{X}$. Bob must then send Charlie a random variable $Y$, taking values $y\in\mathcal{Y}$. Between these two events, Alice may send a communication to Bob, but Bob is not allowed to communicate with Alice. Over the course of many rounds, the $Y$ must verifiably be drawn from a predefined distribution $P(Y|X)$. We assume a sufficient number of rounds are run that Charlie can verify to arbitrary precision that the appropriate distribution is being realised. We assume that there is no shared randomness between the players, that each round must be fully resolved before the next begins, and that Alice and Bob may not maintain any residual correlations between rounds, such that from the perspective of the players each round is essentially an independent instance of the game.

Trivially, the players can always succeed in the task if Alice communicates the corresponding $x$ to Bob each round~\footnote{Alternatively, Alice could sample from $P(Y|x)$ and send the outcome to Bob to forward on to Charlie.}. So, to make the game interesting we will challenge the players to complete the task while minimising the communication from Alice to Bob. Specifically, we will take as the figure of merit the (log-)dimension of the system Alice uses to communicate with Bob each round, with respect to the corresponding theory (i.e., classical, real quantum, or complex quantum). This can be interpreted as the communication cost~\cite{yao1979some, loomis2020optimizing, ding2024quantum}. 

While we use the same axioms to define real and complex quantum theory as Ref.~\cite{renou2021quantum}, we make a departure in that we allow ourselves to assume the dimension of the systems involved. This is because for the practical purposes considered here we care only about the dimensions that we can use to store information; if a system has additional available dimensions we are unaware of, then this is irrelevant if we are not using them to embed information. Correspondingly, the mechanism of our advantage is different: In Ref.~\cite{renou2021quantum}, the departure between real and complex quantum theory can be ascribed to the tensor product composition rule; here, it is due to the different state spaces, which \emph{are} physically distinguishable between the two theories in the presence of constraints on system dimension (quantified according to the respective theories).

\begin{figure}

\includegraphics[width=0.85\linewidth]{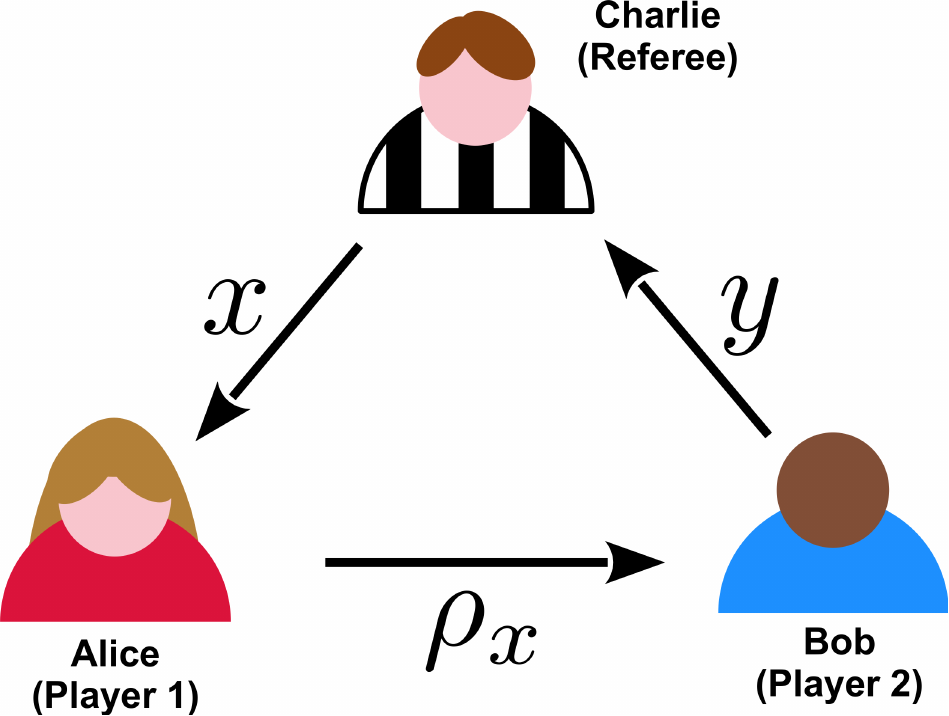}
\caption{{\bf Illustration of the communication game.} In each round the referee Charlie sends Alice a message $x$. Alice then communicates a state $\rho_x$ to Bob, who measures this to produce an output $y$ with probability $R(y|\rho_x)$, which is then sent to Charlie. The aim is for the players to minimise the size of the system Alice communicates to Bob, while ensuring that $y$ is drawn from a target distribution $P(y|x)$.}
\label{fig:game}
\end{figure}

We label the message state (or distribution over states) that Alice sends to Bob upon receipt of $x$ as $\rho_x$. We denote the set of these four states (one for each $x$) by $\rho_{\mathcal{X}}$, and the associated random variable $\rho_X$~\footnote{Note that this also accounts for the scenario where Alice probabilistically selects one of an ensemble of states for each input; in this case, $\rho_x$ denotes the average state of this ensemble.}. For ease of exposition we represent this as a state in complex quantum theory, since real quantum theory and classical theory are recovered as special limits under appropriate constraints. We can express each round as a Markov chain $X\to \rho_X \to Y$~\cite{loomis2020optimizing}. Correspondingly, the communication cost $D$ is given by
\begin{equation}
\label{eq:merit}
D=\log_2\left(\mathrm{dim}[\mathrm{span}(\rho_{\mathcal{X}})]\right).
\end{equation}
Thus, for a given $P(Y|X)$, we seek the smallest $D$ that is sufficient to guarantee success in the task. \figref{fig:game} illustrates a single round of the game.

\subsection{Bargmann Invariants}

To certify the necessity for complex quantum theory, we will use Bargmann invariants~\cite{bargmann1964note, oszmaniec2024measuring, fernandes2024unitary}. For a set of states $\{\rho_n\}$, the Bargmann invariants $\mathfrak{B}$ are defined in terms of the trace of products of the states. An \mbox{$N$-th} order Bargmann invariant consists of products of $N$ states, and we denote the associated states in subscript. For example,
\begin{equation}
\mathfrak{B}_{2,4,5}:=\Tr(\rho_2\rho_4\rho_5).
\end{equation}
These are invariant under rotation of the set of states by any unitary operator $U$, and hence are basis independent. This enables their use as a geometric witness of imaginarity in an ensemble of quantum states. Specifically, if any of the Bargmann invariants are non-real, then this rules out the possibility that there exists \emph{any} basis in which all states in the ensemble can be written with only real elements. In other words, the ensemble must contain imaginarity in every basis, and so requires complex quantum theory. See Appendix \ref{sec:barg} for further details.

\section{Results}

\subsection{Statement of Main Result}

 Here, we will consider a particular version of the game, where $\mathcal{X}=\mathcal{Y}=\{0, 1, 2, 3\}$, i.e., both alphabets are identical, with four elements. The target distribution for the game is $P(Y|X)=(1-\delta_{XY})/3$, where $\delta_{xy}$ is the Kroenecker delta~\cite{boas2006mathematical}, such that in a round where Alice receives $x$, Bob must send Charlie $y\in\mathcal{Y}\backslash x$ with uniform probability. For example, if Alice receives $0$, Bob must send 1, 2, or 3, each with probability $1/3$. 

We will show for this particular version perfect play can be achieved with a cost of $D=1$ for complex quantum theory, i.e., that the states $\rho_{\mathcal{X}}$ can all be embedded within a qubit. We will further show that the cost for approaches constrained to operate within classical theory or real quantum theory must strictly be larger. In other words, while Alice can communicate the requisite information to Bob each round with a qubit, a rebit or bit is not sufficient. 

\subsection{A bit is insufficient}

We first rule out the possibility of a valid strategy with $D=1$ within classical theory. Within classical theory, a two-dimensional system (bit) has two basis states, which we shall label $\{\ket{0},\ket{1}\}$. Alice's communication to Bob each round must be one of these basis states, or a probabilistic mixture thereof.

Let us suppose that upon receipt of $x$, Alice sends Bob $\ket{w}$ with probability $Q(w|x)$, and that upon receipt of $\ket{w}$, Bob sends Charlie $y$ with probability $R(y|w)$. Thus, we must have that $P(Y|X)=R(Y|0)Q(0|X)+R(Y|1)Q(1|X)$. Since $P(Y=x|X=x)=0\forall x$, and the two terms are both non-negative, we must have that both terms vanish when $x=y$. For any given $x$, if both $Q$ components are zero, then Alice sends no communication to Bob. If both $R$ components are zero, then Bob will \emph{never} send $x$ to Charlie, irrespective of the input -- thus rendering him unable to manifest the appropriate distribution for other $x$. Thus, upon receipt of any given $x$, Alice must send Bob one of the basis states with certainty. For $x=0$, let this be $\ket{0}$ without loss of generality. Then, we must have that $R(y|0)=(1-\delta_{0y})/3$.

Let us now consider Alice's action upon receipt of $x=1$. Following the above argument, Alice must send one of the basis states with certainty. Clearly, this cannot be $\ket{0}$, as this yields an incorrect distribution for $Y$ given $x=1$. So, Alice must send $\ket{1}$; thus we require $R(y|1)=(1-\delta_{1y})/3$. Now, consider $x=2$. Again, Alice must send one of the basis states with certainty to Bob. Yet, both states will yield incorrect conditional distributions for $Y$. Thus, it is impossible for Alice to communicate sufficient information to Bob with just a single bit, ruling out a $D=1$ strategy within classical theory.

\subsection{Ruling out mixed quantum state communication}

We next show that if there exists a valid quantum strategy with $D=1$, then it must involve Alice sending pure states to Bob. This follows a similar manner of argument to that for ruling out classical theory above.

Suppose on input $x$, Alice sends mixed state $\rho_x$, with eigenbasis $\{\ket{0}_x,\ket{1}_x\}$ and eigenvalues $\{{\lambda_0}_x,{\lambda_1}_x\}$. Bob measures the received $\rho_x$ to determine the $y$ to be sent to Charlie, obtaining $y$ with probability $R(y|\rho_x)$. From the linearity of quantum theory, it follows that this can be decomposed into $R(y|\rho_x)={\lambda_0}_xR(y|\ket{0}_x)+{\lambda_1}_xR(y|\ket{1}_x)$. As with the classical case, this must vanish for $y=x$; since both terms are non-negative, both terms must vanish. We cannot simultaneously have $R(x|\ket{0}_x)=0$ and $R(x|\ket{1}_x)=0$, as this entails Bob \emph{never} sending $x$ to Charlie for any input, again rendering him unable to manifest the appropriate distribution~\footnote{Note that for this argument it does not matter that the eigenbases for the $\rho_x$ can be different. For a given basis $\{\ket{0},\ket{1}\}$, if $R(y|\ket{0})=R(y|\ket{1})=0$, then for any valid superposition $\alpha\ket{0}+\beta\ket{1}$ we must also have $R(y|\alpha\ket{0}+\beta\ket{1})=0$.}. Thus, we require that for all $\rho_x$ one of the associated ${\lambda}_x$ must be zero -- i.e., all $\rho_x\in\rho_{\mathcal{X}}$ are pure.

\subsection{Valid quantum strategies}

 We now show that valid single-qubit quantum strategies do exist. From the above, we know that this must require that Alice sends Bob pure message states; correspondingly, let us replace the $\rho_x$ with $\ket{\sigma_x}$. Bob then determines an appropriate $y$ by means of a generalised measurement on the state received. Let us invoke Naimark's dilation theorem~\cite{stinespring1955positive}, such that we can view this measurement as a projective measurement on an enlarged space. That is, there exists a unitary operator $U$ such that we can define $\{\ket{\bar{\sigma}_\mathcal{X}}\}$, with $\ket{\bar{\sigma}_x}:=U\ket{\sigma_x}$; generalised measurement operators on the original state space can be mapped to projection operators for subspaces of the dilated space.

These dilated states can be written as
\begin{equation}
\ket{\bar{\sigma}_x}:=\sum_y\sqrt{P(y|x)}\ket{y}\ket{\psi_{xy}}.
\end{equation}
The projection operators are of the form $\Pi_y:=\ket{y}\bra{y}\otimes \mathbb{I}$, such that they measure the first subsystem in this space. The amplitudes ensure the correct probabilities are realised, and the $\{\ket{\psi_{xy}}\}$ are `junk' states resulting from the dilation.  Without loss of generality, any valid strategy can be cast in this form.

The junk states are the only tunable components to the strategy in this form. Moreover, we can immediately make simplifications without loss of generality to the possible overlaps of the message states -- and hence the possible Bargmann invariants of valid strategies. Firstly, note that the terms where $x=y$ have zero amplitude and so the associated $\{\ket{\psi_{xx}}\}$ may be ignored. Secondly, we can consider that for one of the $x$ that the dilation unitary $U$ is configured to set the associated $\ket{\psi_{xy}}$ to $\ket{0}$ without loss of generality. We shall take this to be $x=0$, such that $\ket{\psi_{0y}}=\ket{0}\forall y$. Since there is no $y=0$ term for $x=0$, we can also set $\ket{\psi_{10}}=\ket{0}$.

Since we are assuming the four message states can be encoded into a qubit, they must be linearly dependent. Specifically, since $\ket{\sigma_0}\neq\ket{\sigma_1}$, there must exist $\{\alpha_0,\alpha_1\}$ and $\{\beta_0,\beta_1\}$ such that
\begin{align}
\label{eq:lindep}
\ket{\sigma_2}&=\alpha_0\ket{\sigma_0}+\alpha_1\ket{\sigma_1}\nonumber\\
\ket{\sigma_3}&=\beta_0\ket{\sigma_0}+\beta_1\ket{\sigma_1}.
\end{align}
Since the dilated message states are related to the originals by a unitary operator, they must share the same pairwise overlaps and linear dependencies. That is, the relations in Eq.~\eqref{eq:lindep} also hold for the dilated states.

By inspecting the $P(0|x)$ and $P(1|x)$ terms, it can be immediately deduced that we must have $|\alpha_0|=|\alpha_1|=|\beta_0|=|\beta_1|=1$ in order to satisfy the correct distributions. Since an overall phase factor in front of message state is irrelevant, we can therefore set $\alpha_0=\beta_0=1$ without loss of generality. Note that this will affect the pairwise overlaps between the states, scaling them by a phase factor; crucially though, it does \emph{not} affect the Bargmann invariants.

To satisfy $P(2|2)=P(3|3)=0$, we must also have that $\alpha_0\ket{0}+\alpha_1\ket{\psi_{12}}=0$ and $\beta_0\ket{0}+\beta_1\ket{\psi_{13}}=0$. Together with the above constraint on the magnitude of the scalar parameters, these can only be satsfied if $\ket{\psi_{12}}$ and $\ket{\psi_{13}}$ are related to $\ket{0}$ by a scalar phase factor. Correspondingly, we replace these by $e^{i\varphi}$ and $e^{i\phi}$ respectively, and drop the $\ket{0}$ representing the (now trivial) second subsystem in the dilated states. Moreover, we can use the freedom to scale $\ket{\bar{\sigma}_1}$ by a phase factor to set $\varphi=0$. From the above conditions, we now have that $1+\alpha_1=0$ and $1+\beta_1e^{i\phi}=0$; thus, $\alpha_1=-1$ and $\beta_1=-e^{-i\phi}$.

Finally, in order to satisfy $P(2|3)=P(3|2)=1/3$, we must have that $|1-e^{i\phi}|=1$ and $|1-e^{-i\phi}|=1$. These are (only) satisfied by $\phi=\pm\pi/3$. Correspondingly, we obtain two valid sets of dilated message states, depending on whether we take the positive or negative solution for $\phi$. The associated dilated message states are given by
\begin{align}
\ket{\bar{\sigma}_0}&=\frac{1}{\sqrt{3}}(\ket{1}+\ket{2}+\ket{3})\nonumber\\
\ket{\bar{\sigma}_1}&=\frac{1}{\sqrt{3}}(\ket{0}+\ket{2}+e^{\pm \frac{i\pi}{3}}\ket{3})\nonumber\\
\ket{\bar{\sigma}_2}&=\frac{1}{\sqrt{3}}(-\ket{0}+\ket{1}+e^{\mp \frac{i\pi}{3}}\ket{3})\nonumber\\
\ket{\bar{\sigma}_3}&=\frac{1}{\sqrt{3}}(-e^{\mp \frac{i\pi}{3}}\ket{0}+\ket{1}+e^{\pm \frac{i\pi}{3}}\ket{2}).
\end{align}
It can readily be verified that these lead to the correct $P(Y|X)$ being generated, and sit within a two-dimensional space. They hence present valid strategies with $D=1$. Moreover, as we have not ceded any generality in the process of deducing these strategies in terms of Bargmann invariants, \emph{any} valid $D=1$ strategy must have message states with the same set of Bargmann invariants as these dilated message states.

\subsection{Valid strategies require complex quantum theory}

Our final step is to show that the valid strategies above -- the only valid $D=1$ strategies within quantum theory -- require complex quantum theory and cannot be realised with real quantum theory. To do this we calculate the 3rd-order Bargmann invariants of the message states. In fact, it is sufficient to show that a single one of these is non-real in order to rule out real quantum theory. Considering $\mathfrak{B}_{0,1,2}$ explicitly:
\begin{align}
\mathfrak{B}_{0,1,2}&=\braket{\sigma_0}{\sigma_1}\braket{\sigma_1}{\sigma_2}\braket{\sigma_2}{\sigma_0}=\braket{\bar{\sigma}_0}{\bar{\sigma}_1}\braket{\bar{\sigma}_1}{\bar{\sigma}_2}\braket{\bar{\sigma}_2}{\bar{\sigma}_0}\nonumber\\
&=\frac{1}{27}\left((1+e^{\pm \frac{i\pi}{3}})\times(e^{\mp \frac{2i\pi}{3}}-1)\times(1+e^{\pm \frac{i\pi}{3}})\right)\nonumber\\
&=\mp\frac{i}{3\sqrt{3}}.
\end{align}
That this is imaginary is sufficient to verify that these require complex quantum theory. In fact, all non-trivial 3rd-order Bargmann invariants of these sets of states are imaginary (see Appendix \ref{sec:overlaps}). Thus, for this task a $D=1$ strategy is only viable within complex quantum theory, while real quantum theory is insufficient to achieve this minimal cost. We thus have a strict advantage demonstrated by complex quantum theory in this task.

\section{Discussion}

By considering a conceptually straightforward communication task, and leveraging the geometry of quantum states, we have revealed an example of where complex quantum theory can exhibit strict operational advantages over real quantum theory. This adds substantive evidence that complex amplitudes can indeed provide a meaningful resource in quantum information processing, and that the constraints of real quantum theory -- as defined in Ref.~\cite{renou2021quantum} -- demonstrably limit information processing power compared to complex quantum theory. The basis-independent nature of our complexity witness ensures that this result is not an artefact of the choice of basis for the system, and since the particular task considered here requires only a single qubit, it is a readily-testable proposal.

Though a seemingly rather simple result, it has potentially significant consequences, and raises a number of lines of inquiry. Firstly, the robustness of the result. This can be interpreted in two closely-related manners: One, the smallest error Alice and Bob can achieve with real quantum theory if they are allowed to imperfectly approximate $P(Y|X)$; the other, the sensitivity of the result to noise in the channel. For example, in the case of the former we could parameterise this as the average (over inputs $\mathcal{X}$) distance between the realised and target distributions. For the trace distance, this would be $d=\frac{1}{|\mathcal{X}|}\sum_{x\in\mathcal{X}}\frac{1}{2}\sum_{y\in\mathcal{Y}}|P(y|x)-\tilde{P}(y|x)|$, where $P$ and $\tilde{P}$ are the target and realised distributions respectively. We strongly conjecture that the smallest achievable distance with a rebit strategy is finite. Through an explicit construction we show that $d=1/12$ is achievable for a rebit strategy. Details are given in Appendix \ref{sec:bounds}. Based on a rudimentary attempt at optimisation, we weakly conjecture that this is the optimal rebit strategy according to the average trace distance, but leave a more thorough examination for future work.

Secondly, here we have witnessed that complex quantum theory is necessary for the task, but have prescribed no measure of how much imaginarity or complexity is needed. To take an ensemble-based resource theory of imaginarity or complexity further, such a measure will be required; this measure could also then be applied to understand complexity as a resource in quantum information processing more generally. A third line of inquiry takes a more philosophical mien: given a `genuinely complex' qubit requires a two rebit embedding to replicate with real quantum theory, could the above task be used to benchmark whether a two-dimensional quantum system is a genuinely complex qubit, and within the composite rebit real quantum theory interpretation of the task to what extent can the underlying rebits forming the qubit be individually addressed? 

Finally, with complexity indeed a meaningful quantum resource, can we identify imaginarity or complexity as the actual resource underlying certain quantum algorithms or information processing tasks? In this particular task, we do not expect a significant scaling advantage, irrespective of the distribution $P(Y|X)$~\footnote{Specifically, an $n_C$-dimensional complex quantum system can be represented by $N_C=n_c^2-1$ generalised Bloch vectors, while an $n_R$-dimensional real quantum system has $N_R=(n_r+2)(n_r-1)/2$. Asymptotically then, $N_C=N_R$ for $n_R\approx \sqrt{2}n_C$. So we expect at most a roughly constant factor scaling difference between real and complex quantum theory for this task.}. Nevertheless, the class of task considered here readily generalises to more important and impactful applications that may show more significant scaling advantages, particularly when memory and/or more players are involved. The general setting is typical of communication tasks, with the cost representing the communication complexity. Variants with a greater number of parties, and multi-directional communication would describe communication networks proper.

Moreover, our result is of particular relevance for stochastic simulation. Consider, rather than a communication between parties, the communicated system instead interpreted as a memory that is updated and preserved between rounds, and used to sequentially generate a stochastic process. This is stochastic simulation, and our task represents a particular instance of this. In this context, a quantum advantage in memory cost has been demonstrated, even when the simulated process is classical~\cite{gu2012quantum, aghamohammadi2018extreme, elliott2021memory}. Further, this quantum advantage has been enhanced by using complex amplitudes to achieve quantum dimension reduction~\cite{liu2019optimal} -- in some cases with significant effect~\cite{elliott2020extreme} -- albeit without explicitly ruling out the possibility of an equivalent real quantum theory approach under a suitable basis transformation. These works show an unbounded gap between classical and quantum theory approaches to the task. Our work shows that complexity is indeed a key resource in maximising quantum advantage in this context; further exploration of our work in this broader context is a promising future avenue. Such scenarios also readily generalise to input-dependent processes such as adaptive agents~\cite{elliott2022quantum}, suggesting complexity could also be a resource for realising quantum-enhanced artificial intelligence.

\acknowledgments
This work was funded by the University of Manchester Dame Kathleen Ollerenshaw Fellowship. We thank Sion Meredith for assistance with Bayesian optimisation.

\appendix

\section{Real and Complex Quantum Theory}
\label{sec:rcqt}

We follow Ref.~\cite{renou2021quantum} in our definitions of real and complex quantum theory. Quantum theory is defined from the following axioms:
\begin{itemize}
\item {\bf State space.} Any physical system $S$ has a corresponding Hilbert space $\mathcal{H}_S$. The system is characterised by its density operator $\rho$, a positive semidefinite, Hermitian, unit trace operator acting on $\mathcal{H}_S$.
\item {\bf Measurement.} A measurement $M$ on $S$ corresponds to a set of positive semidefinite Hermitian operators $\{M_x\}$ acting on $\mathcal{H}_S$ that satisfy $\sum_x M_x=\mathbb{I}$, with $x$ indicating the outcome of the measurement.
\item {\bf Born Rule.} If $S$ is described by a density operator $\rho$, the probability of obtaining outcome $x$ from measurement $M$ is given by $P(x)=\Tr(\rho M_x)$.
\item {\bf Composition.} A composite of systems $S$ and $T$ with Hilbert spaces $\mathcal{H}_S$ and $\mathcal{H}_T$ respectively has a corresponding Hilbert space $\mathcal{H}_{ST}=\mathcal{H}_S\otimes\mathcal{H}_T$. Operators that describe measurements or transformations of one system act trivially on the space associated to the other. The state representing an independent preparation of the two systems is the tensor product of these two preparations.
\end{itemize}
The nature of the Hilbert space associated to the system defines the type of quantum theory. Specifically, in complex quantum theory these are complex Hilbert spaces; in real quantum theory they are real Hilbert spaces.

\section{Bargmann Invariants}
\label{sec:barg}

Recall from the main text that the Bargmann invariants~\cite{bargmann1964note} of a set of quantum states $\{\rho_n\}$ are defined as the trace of products of the states. It was stated that they are invariant under unitary rotation of the states, i.e., that the Bargmann invariants of $\{\rho_n\}$ and $\{U\rho_nU^\dagger\}$ are identical for any unitary operator $U$. This follows straightforwardly from the cyclicity of the trace and that $U^\dagger U=\mathbb{I}$. For example, 
\begin{equation}
\Tr(U\rho_2U^\dagger U\rho_4U^\dagger U\rho_5U^\dagger) = \Tr(\rho_2\rho_4\rho_5).
\end{equation}
Since unitary rotations of the set of states is essentially equivalent to changing the basis of the states, this substantiates that Bargmann invariants are basis-invariant. Due to the cyclicity of the trace, Bargmann invariants are also unchanged over cyclic permutation of the labels, e.g., $\mathfrak{B}_{2,4,5}=\mathfrak{B}_{4,5,2}=\mathfrak{B}_{5,2,4}$. The set of values taken by Bargmann invariants of any given order is also independent of the label ordering of the states.

The resource theory of imaginarity~\cite{wu2021operational} identifies imaginarity in a state with imaginary elements in the density matrix of the state. Correspondingly, if an ensemble of states all possess no imaginarity in a given basis, then all density matrix elements of the states in this basis must be real. It can readily be seen that this requires that all Bargmann invariants, calculated in this basis, must be real. However, since the Bargmann invariants are basis independent, it does not matter which choice of basis we use to calculate them -- if there exists a basis choice where they can be expressed with real amplitudes, all Bargmann invariants must be real. Conversely, if any of the Bargmann invariants are non-real, then there cannot exist a basis in which all states in the ensemble possess no imaginarity. Thus, complex Bargmann invariants are sufficient to witness basis-independent imaginarity, and hence the need for complex quantum theory to describe the ensemble~\cite{oszmaniec2024measuring}. Note that first and second order Bargmann invariants are insufficient to witness imaginarity -- these are always real and non-negative. We must turn to third order or higher. Since this stricter notion of imaginarity is defined only with respect to an ensemble of states, it has been termed ``set imaginarity"~\cite{fernandes2024unitary}.

When all states in the ensemble are pure, then the Bargmann invariants can be expressed as cyclic products of the state overlaps~\cite{oszmaniec2024measuring, fernandes2024unitary}. That is, for an ensemble of states $\{\ket{\psi_n}\}$, then, for example,
\begin{equation}
\mathfrak{B}_{2,4,5}:=\braket{\psi_2}{\psi_4}\braket{\psi_4}{\psi_5}\braket{\psi_5}{\psi_2}.
\end{equation}
In addition to being basis invariant, for pure states the Bargmann invariants are also unchanged under scalar phase multiplication of any of the states.

\section{Message state overlaps and Bargmann invariants}
\label{sec:overlaps}

In Table \ref{tab:overlap} we specify the full set of overlaps between the message states for the valid $D=1$ quantum strategies, from which all the Bargmann invariants can readily be calculated.

\begin{table}
\centering
\caption{Overlap of the message states $\braket{\sigma_a}{\sigma_b}$ for valid quantum strategies with $D=1$.}
\begin{tabular}{>{\centering\hspace{0pt}} m{0.08\linewidth}|>{\centering\hspace{0pt}}m{0.11\linewidth}>{\centering\hspace{0pt}}m{0.11\linewidth}>{\centering\hspace{0pt}}m{0.11\linewidth}>{\centering\hspace{0pt}}m{0.11\linewidth}>{\hspace{0pt}}m{0.00001\linewidth}}
 {\backslashbox[0pt][l]{$\hspace{0.25em}a\hspace{0.25em}$}{$b\hspace{0.25em}$}} & 0 & 1 & 2 &  3 &   \\ 
\cline{1-5}
0 &  1 & $\dfrac{e^{\pm\frac{i\pi}{6}}}{\sqrt{3}}$  & $\dfrac{e^{\mp\frac{i\pi}{6}}}{\sqrt{3}}$  & $\dfrac{e^{\pm\frac{i\pi}{6}}}{\sqrt{3}}$ & \vspace{0.7em} \\
1 & $\dfrac{e^{\mp\frac{i\pi}{6}}}{\sqrt{3}}$  &  1 & $\dfrac{e^{\mp\frac{5i\pi}{6}}}{\sqrt{3}}$  & $\pm\dfrac{i}{\sqrt{3}}$  & \vspace{0.7em} \\
2 & $\dfrac{e^{\pm\frac{i\pi}{6}}}{\sqrt{3}}$  &  $\dfrac{e^{\pm\frac{5i\pi}{6}}}{\sqrt{3}}$ & 1  & $\dfrac{e^{\mp\frac{i\pi}{6}}}{\sqrt{3}}$ & \vspace{0.7em} \\
3 & $\dfrac{e^{\mp\frac{i\pi}{6}}}{\sqrt{3}}$ & $\mp\dfrac{i}{\sqrt{3}}$  &  $\dfrac{e^{\pm\frac{i\pi}{6}}}{\sqrt{3}}$ &   1 & \vspace{0.7em}
\end{tabular}
\label{tab:overlap}
\end{table}

From these overlaps, we can deduce an appropriate embedding of the dilated message states into states of a qubit using a reverse Gram-Schmidt procedure~\cite{dennery1996mathematics}. A valid embedding is as follows:
\begin{align}
\ket{\sigma_0}&=\ket{0}\nonumber\\
\ket{\sigma_1}&=\frac{e^{\pm\frac{i\pi}{6}}}{\sqrt{3}}\ket{0}+\frac{\sqrt{2}}{\sqrt{3}}\ket{1}\nonumber\\
\ket{\sigma_2}&=\frac{e^{\mp\frac{i\pi}{6}}}{\sqrt{3}}\ket{0}-\frac{\sqrt{2}}{\sqrt{3}}\ket{1}\nonumber\\
\ket{\sigma_3}&=\frac{e^{\pm\frac{i\pi}{6}}}{\sqrt{3}}\ket{0}+\frac{\sqrt{2}}{\sqrt{3}}e^{\pm\frac{2i\pi}{3}}\ket{1}.
\label{eq:qubitstates}
\end{align}

\begin{figure}
\includegraphics[width=\linewidth]{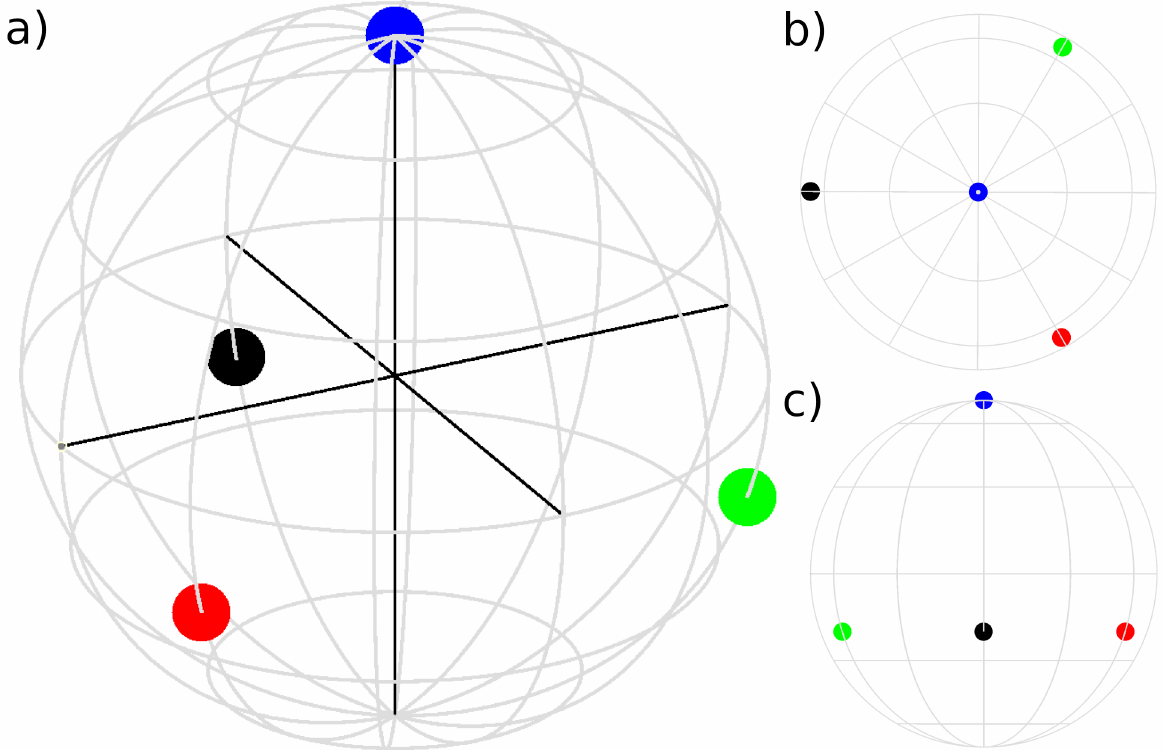}
\caption{{\bf Message states for the qubit strategy.} (a) Bloch sphere plot of the message states from Eq.~\eqref{eq:qubitstates}, for the $\phi=\pi/3$ solution (the $\phi=-\pi/3$ solution differs only in a mirror symmetry). Also shown (b) looking down the $z$-axis and (c) along the $y$-axis.}
\label{fig:states}
\end{figure}

We plot these states in \figref{fig:states}. It is clear that no three message states -- let alone all four -- sit within a great circle of the Bloch sphere, a necessary and sufficient condition for a rebit strategy. Evidently, infinitesimal perturbations cannot be used to constrain all four states to a single great circle. Since the measurement probabilities will vary smoothly with changes to the states and measurement operators, we therefore conjecture that any rebit strategy must be a finite distance from being valid.

Moreover, they provide a direct route to straightforwardly calculating any Bargmann invariant, simply by multiplying the appropriate overlaps together. In the main text we showed that $\mathfrak{B}_{0,1,2}=\mp i/3\sqrt{3}$ and therefore that there is no consistent basis in which the four message states can be encoded with only real amplitudes. In fact, all of the non-trivial (i.e., no repeated index) 3rd-order Bargmann invariants are imaginary. Specifically, there are (up to cyclic permutations) 8 such invariants, given by
\begin{align}
\mathfrak{B}_{0,1,2}&=\mathfrak{B}_{0,2,3}=\mathfrak{B}_{0,3,1}=\mathfrak{B}_{1,3,2}=\mp\frac{i}{3\sqrt{3}}\nonumber\\
\mathfrak{B}_{0,2,1}&=\mathfrak{B}_{0,3,2}=\mathfrak{B}_{0,1,3}=\mathfrak{B}_{1,2,3}=\pm\frac{i}{3\sqrt{3}}.
\end{align}

\section{Bounding the distance for rebit strategies}
\label{sec:bounds}

As noted in the Main Text, one way to quantify the error is via the trace distance over all inputs, i.e., $d=\frac{1}{|\mathcal{X}|}\sum_{x\in\mathcal{X}}\frac{1}{2}\sum_{y\in\mathcal{Y}}|P(y|x)-\tilde{P}(y|x)|$.

We can readily deduce the smallest such error for a communication-free (i.e., $D=0$) strategy, wherein Bob simply samples from $P(Y)$. Assuming a uniform distribution over inputs $x\in\mathcal{X}$, for the particular version here this leads to $d=1/4$. For this strategy, in each round there is a 1/4 chance for Charlie to immediately deduce that Alice and Bob are not producing the correct distribution; this occurs when Bob sends a $y$ that is the same as the $x$ sent to Alice.

For a rebit strategy, we are able to upper-bound the smallest error achievable at $d\leq1/12$, through explicit construction of a strategy that achieves this value. Consider the following trio of dilated states for three of the inputs:
\begin{align}
\ket{\bar{\sigma}_0}&=\frac{1}{\sqrt{2}}(\ket{1}+\ket{2})\nonumber\\
\ket{\bar{\sigma}_1}&=\frac{1}{\sqrt{2}}(\ket{0}-\ket{2})\nonumber\\
\ket{\bar{\sigma}_2}&=\frac{1}{\sqrt{2}}(\ket{0}+\ket{1})
\end{align}
It can be seen that $\ket{\bar{\sigma}_0}+\ket{\bar{\sigma}_1}=\ket{\bar{\sigma}_2}$, and so these three states can be embedded within a qubit. Moreover, since all amplitudes are real, they can be embedded within a rebit. Let us also set $\ket{\bar{\sigma}_3}=\ket{\bar{\sigma}_2}$, such that all four states can be embedded within a rebit. Bob produces statistics as follows:
\begin{enumerate}
\item Bob measures the (dilated) state he receives in the $\{\ket{0},\ket{1},\ket{2}\}$ basis.
\item Bob generates a random number $r$ in the interval $[0,1]$. If $r<2/3$, he sends $y$ equal to his measurement outcome.  Otherwise, he sends $y=3$.
\end{enumerate}
It can be verified that this will generate the correct distribution for $x=0$, $1$, or $2$. However, for $x=3$ it will instead generate the statistics that would have been required for $x=2$. This leads to $d=1/12$. For this strategy, on one of the inputs ($x=3$) Charlie will immediately be able to deduce that the generated distribution is incorrect with 1/3 chance (i.e., when Bob sends $y=3$). With a slightly modified strategy (where Bob has a further post-processing step in which instead of sending $y=2$ or $y=3$ when instructed to according to the above protocol, he randomly chooses to send one of them with equal probability) this can be altered to Charlie having a 1/6 chance of immediately deducing that the distribution is incorrect, but now for two of the inputs; this modified strategy still has $d=1/12$.

Through a rudimentary search over rebit strategies with gradient descent and Bayesian optimisation approaches, we were unable to find a rebit strategy with smaller error than this. However, it should be noted that the cost function landscape (i.e., $d$) is beset with local minima. We therefore only weakly conjecture that $d=1/12$ the the minimum distance for a rebit strategy.

\bibliography{ref}

\end{document}